\documentclass[12pt,twoside]{article}
\usepackage[mathscr]{eucal}
\usepackage{amsmath,amsfonts,amssymb,amsthm,mathabx,empheq}
\usepackage{times}
\usepackage{pdfsync}
\usepackage{cite}
\usepackage{url}
\usepackage{hyperref}
\usepackage{tensor}
\usepackage{color}
\usepackage{multicol}
\usepackage{bbold}

\advance\textheight\topskip
\allowdisplaybreaks[1]

\newcommand\td{\text{d}}
\newcommand\cO{{\cal O}}

\newcommand{\p}{\partial}
\newcommand{\be}{\begin{equation}}
\newcommand{\ee}{\end{equation}}
\newcommand{\bea}{\begin{eqnarray}}
\newcommand{\eea}{\end{eqnarray}}
\def\nn{\nonumber}
\def\bz{\bar z}

\def\nb{\bar\nabla}
\def\barg{\bar g}
\def\ga{\gamma_{z\bz}}

\newcommand*\xbar[1]{%
  \hbox{%
    \vbox{%
      \hrule height 0.5pt 
      \kern0.3ex
      \hbox{%
        \kern-0.0em
        \ensuremath{#1}%
        \kern-0.0em
      }%
    }%
  }%
}

\hypersetup{
colorlinks=true,
linktoc=page,    
linkcolor=blue,
citecolor=blue,
urlcolor=blue,
colorlinks=true,
}

\DeclareFontFamily{OT1}{rsfs}{} \DeclareFontShape{OT1}{rsfs}{m}{n}{
<-7> rsfs5 <7-10> rsfs7 <10-> rsfs10}{}
\DeclareMathAlphabet{\mycal}{OT1}{rsfs}{m}{n}

\begin{document}
\title{Linearized gravity and soft graviton theorem in de Sitter spacetime}

\author{Pujian Mao and Bochen Zhou}

\date{}

\def\mytitle{Linearized gravity and soft graviton theorem in de Sitter spacetime}

\addtolength{\headsep}{4pt}

\begin{centering}

  \vspace{1cm}

  \textbf{\Large{\mytitle}}

  \vspace{1.5cm}

  {\large Pujian Mao and Bochen Zhou}

\vspace{.5cm}

\vspace{.5cm}
\begin{minipage}{.9\textwidth}\small \it  \begin{center}
    Center for Joint Quantum Studies and Department of Physics,\\
     School of Science, Tianjin University, 135 Yaguan Road, Tianjin 300350, China
 \end{center}
\end{minipage}

\end{centering}


\vspace{1cm}

\begin{center}
\begin{minipage}{.9\textwidth}
\textsc{Abstract}. We study the linearized gravity theory in the Newman-Unti gauge in the near horizon region of the de Sitter spacetime. The linearized Einstein equation involves the cosmological constant. The near horizon symmetry consists of near horizon supertranslation and near horizon superrotation. We compute the near horizon supertranslation charge and find the proper near horizon fall-off conditions which uncover a soft graviton theorem from the Ward identity of the near horizon supertranslation. 
\end{minipage}
\end{center}

\begin{center}
Emails: pjmao@tju.edu.cn,\, zhoubch@tju.edu.cn
\end{center}

\thispagestyle{empty}

\newpage
\tableofcontents

\section{Introduction}

Black holes are recently shown to carry soft hairs \cite{Hawking:2016msc} which reveals a much richer structure that black holes can have. The new degrees of freedom are labeled by the near horizon symmetries \cite{Donnay:2015abr,Averin:2016ybl,Afshar:2016wfy,Setare:2016jba,Mao:2016pwq,Setare:2016vhy,Afshar:2016uax,Grumiller:2016kcp,Donnay:2016ejv,Setare:2016msj,Sheikh-Jabbari:2016npa,Cai:2016idg,Hawking:2016sgy,Afshar:2016kjj,Shi:2016jtn,Eling:2016qvx,Akhmedov:2017ftb,Grumiller:2019tyl,Grumiller:2019fmp,Adami:2020amw,Grumiller:2020vvv,Donnay:2020yxw,Adami:2020ugu,Adami:2021sko,Adami:2021nnf,Adami:2021kvx,Liu:2022uox,Adami:2022ktn,Taghiloo:2022hxc,Mao:2022ldv,Aggarwal:2023qwl}. Black hole horizons can be considered as an inner boundary of the spacetime which share many common features as the null infinity, see, e.g., investigations in \cite{Ashtekar:2024mme,Ashtekar:2024bpi}. At null infinity, a triangle equivalence was proven \cite{Strominger:2017zoo}, which connects asymptotic symmetry, memory effect, and soft theorem. Inspired by the null infinity triangle relation, the black hole memory effect was subsequently proposed \cite{Donnay:2018ckb,Rahman:2019bmk,Bhattacharjee:2020vfb,Adami:2021nnf} and it is closely related to the near horizon symmetry. Another branch following the triangle relation is the soft theorems relevant to near horizon symmetry which have been investigated for the Schwarzschild black hole \cite{Cheng:2022xyr,Cheng:2022xgm,Mao:2023dsy}. Nevertheless, horizons are not exclusive to black holes. The expanding universe leads to a cosmological horizon. A cosmological horizon is locally very
similar to the black hole horizon in the sense that they are both codimension one null hypersurface which cut spacetime into two parts. They can be considered as the casual boundary of local observer and the physically relevant investigations normally reside at only one side of the horizons. Soft theorems of gauge theory in de Sitter (dS) spacetime are obtained from the near horizon symmetry \cite{Mao:2023rca}. The computations in \cite{Mao:2023rca} are somewhat similar to the Schwarzschild case \cite{Cheng:2022xyr,Cheng:2022xgm}, while the physical motivation and consequence are very different. The cosmological horizon is the outer boundary of the inside observer. Because the dS cosmological universe is expanding so fast, there are events which will never be seen by an observer inside. Considering the fact that we do live in an expanding universe, the near cosmological horizon analysis can be intuitively understood as the real world asymptotic analysis near null infinity. Technically, there is well defined flat limit from the cosmological solution
\cite{Barnich:2012aw,Compere:2019bua,Compere:2020lrt}. The cosmological
horizon becomes the null infinity as the cosmological constant tends to zero. Correspondingly, the near horizon
soft theorems derived in dS spacetime should recover the flat spacetime soft theorem in the flat limit. This is different from the black hole case. For instance, one can recover the Minkowski spacetime when the mass parameter of the Schwarzschild black hole is zero. But the horizon of the black hole just disappears in this limit. The aim of the present work is to extend the previous study \cite{Mao:2023rca} in dS spacetime to linearized gravity.

For the linearization of Einstein theory about dS spacetime, the linearized equations of motion involve the cosmological constant (see, e.g., in  \cite{deVega:1998ia,Ashtekar:2015lla,Date:2015kma,Chu:2016qxp,Compere:2023ktn}) to incorporate with gauge invariance reduced from the diffeomorphism invariance of the Einstein theory. The effect of the cosmological constant is well understood in both the asymptotic and near horizon analysis. But the relevance to a soft theorem has only been considered as a parameter in the metric of the background spacetime \cite{Mao:2023rca}. Since the cosmological constant modifies the equations of motion, it is very questionable that if the nice structure revealed in \cite{Mao:2023dsy} or \cite{Mao:2023rca} can be derived for linearized gravity in dS spacetime. The potential effect of the cosmological constant in the equations of motion is the main extension of this work to previous studies \cite{Mao:2023dsy,Mao:2023rca}.

In this work, we apply the Newman-Unti (NU) gauge \cite{Newman:1962cia} for the linearized gravity theory in dS spacetime. In the near cosmological horizon region, we impose traceless fall-off conditions. We compute the near horizon symmetry which consists of near horizon supertranslation and near horizon superrotation. The near horizon solution space is specified and the supertranslation charge is derived. We find that there is an interesting reduction of the near horizon solution space which is invariant under supertranslation and leads to a natural split of the supertranslation charge into soft and hard parts. We found such a configuration for the Schwarzschild black hole case in a previous work \cite{Mao:2023dsy} and considered that as a coincidence. Now such configuration also arises for the dS spacetime. This may suggest that there is a common structure in the near horizon region which is subject to a soft theorem. A soft graviton theorem in coordinate space is derived from the Ward identity of the near horizon supertranslation in the reduced solution space. After transforming the soft graviton theorem into the momentum space, there is a natural flat limit and it recovers the flat space soft graviton theorem.   

This paper is organized as follows. In the next section, we present the linearized Einstein equation in dS spacetime, compute the near horizon symmetry and present the solution space of the linearized theory. We specify a reduction of the solution space where the supertranslation charge can split into a soft piece and a hard piece. In section \ref{soft}, a soft graviton theorem is derived from the Ward identity of the near horizon supertranslation charge. The soft theorem has a desired flat limit in the momentum space. The last section is devoted to conclusion and discussion. There is one Appendix which presents the details of the modification of the stress tensor that coupled to the gravity theory.


\section{Near horizon symmetries and charges}

In this section, we will study the linearized Einstein theory in the dS spacetime. The NU gauge \cite{Newman:1962cia} (see also \cite{Barnich:2011ty,Conde:2016rom}) will be adapted into the linearized theory. We will perform the standard near horizon analysis. The near horizon symmetry, solution space, and surface charge will be computed.

\subsection{Near horizon form of the dS spacetime in NU gauge}

\begin{figure}[http]
    \centering
\includegraphics[width=0.4\linewidth]{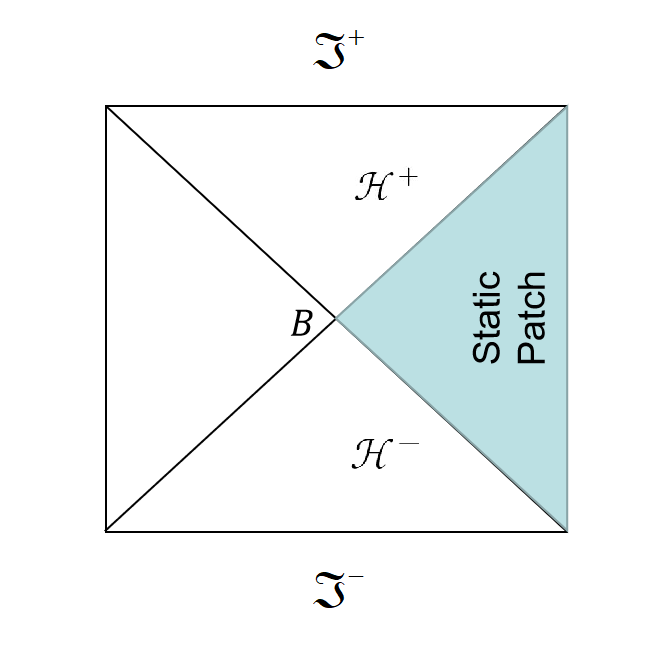}
\caption{Penrose diagram of de Sitter spacetime}
\label{f1}
\end{figure}

We start from the static patch in dS spacetime, which is part of the full dS spacetime as shown in Fig. \ref{f1}. 
In static coordinates $(t,r,z,\bar{z})$, the line element of the dS spacetime is
\begin{align}
    &\mathrm{d}s^2=-F(r)\mathrm{d}t^2+F(r)^{-1}\mathrm{d}r^2+2\Omega(r)^2\gamma_{z\bar{z}}\mathrm{d}z\mathrm{d}\bar{z}\, , \\
    &F(r)=\frac{2 r}{\ell}-\frac{r^2}{\ell^2}\, , \qquad \Omega=\ell-r\, , \qquad \gamma_{z\bar{z}}=\frac{2}{(1+z\bar{z})^2}\, ,
\end{align}
where $\ell$ is the dS radius and it is related to the cosmological constant by $\Lambda=\frac{3}{\ell^2}$. Note that we have set the cosmological horizon at $r=0$. By introducing the retarded time coordinate $u=t-\frac{1}{2} \ell \log\left(\frac{r}{2\ell -r}\right)$, the line element can be written as
\begin{equation}\label{dS}
 \mathrm{d}s^2=\barg_{\mu\nu}\td x^\mu \td x^\nu = -F(r)\mathrm{d}u^2-2\mathrm{d}r\mathrm{d}u+2\Omega(r)^2\gamma_{z\bar{z}}\mathrm{d}z\mathrm{d}\bar{z}\, ,
\end{equation}
which covers only the $\mathcal{H}^{-}$ part of the horizon. A similar analysis could be performed for $\mathcal{H}^{+}$ simply by a time reverse transformation of dS spacetime near the bifurcation sphere $B$. In the rest of this paper, we just focus on the $\mathcal{H}^{-}$ part.

\subsection{Linearization in dS spacetime}

We linearize Einstein theory in dS spacetime \eqref{dS}. The metric expands as $g_{\mu\nu}=\barg_{\mu\nu}  +  \kappa h_{\mu\nu}$. The inverse metric is $g^{\mu\nu}=\barg^{\mu\nu}  -  \kappa h^{\mu\nu} + \cO(\kappa^2)$, where the indices are now raised by $\barg^{\mu\nu}$. Up to the first correction order $\cO(\kappa)$, the connection is given by
\be
\Gamma^\alpha_{\mu\nu}={\bar\Gamma}^\alpha_{\mu\nu}+\frac{\kappa}{2} \barg^{\alpha\beta}\left(\nb_\mu h_{\nu\beta} + \nb_\nu h_{\mu\beta} - \nb_\beta h_{\mu\nu} \right).
\ee
We define
\be
C^\alpha_{\mu\nu}=\barg^{\alpha\beta}\left(\nb_\mu h_{\nu\beta} + \nb_\nu h_{\mu\beta} - \nb_\beta h_{\mu\nu} \right),
\ee
which is very useful for the computation of the curvature tensor,
\begin{align}
{R_{\mu\nu\alpha}}^\beta&=\p_\nu \Gamma^\beta_{\mu\alpha} - \p_\mu \Gamma^\beta_{\nu\alpha} + \Gamma^\tau_{\mu\alpha} \Gamma^\beta_{\nu\tau} - \Gamma^\tau_{\nu\alpha}\Gamma^\beta_{\mu\tau}\nn\\
&={\bar R_{\mu\nu\alpha} } ^{\quad\,\,\,\beta} + \frac{\kappa}{2}(\nb_\nu C^\beta_{\mu\alpha}-\nb_\mu C_{\nu\alpha}^\beta) + \cO(\kappa^2)\\
&={\bar R_{\mu\nu\alpha} } ^{\quad\,\,\,\beta} + \frac{\kappa}{2} \nb_\nu \left(\nb_\mu h^\beta_\alpha + \nb_\alpha h_\mu^\beta - \nb^\beta h_{\mu\alpha}\right)\nn \\
&\hspace{2cm}- \frac{\kappa}{2} \nb_\mu \left(\nb_\nu h^\beta_\alpha +  \nb_\alpha h_\nu^\beta - \nb^\beta h_{\nu\alpha}\right) + \cO(\kappa^2).
\end{align}
The Ricci tensor is defined from the curvature tensor as
\be
R_{\mu\alpha}={R_{\mu\nu\alpha}}^\nu
=\bar R_{\mu\alpha}  + \frac{\kappa}{2} \nb_\nu \left(\nb_\mu h^\nu_\alpha + \nb_\alpha h_\mu^\nu - \nb^\nu h_{\mu\alpha}\right) 
- \frac{\kappa}{2} \nb_\mu  \nb_\alpha h_\nu^\nu  + \cO(\kappa^2).
\ee
Finally, the Ricci scalar is obtained as
\be
R=g^{\mu\alpha}R_{\mu\alpha}=\bar R - \kappa h^{\mu\alpha}\bar R_{\mu\alpha}+ \frac{\kappa}{2} \nb_\nu \left(\nb_\mu h^{\nu\mu} + \nb^\mu h_\mu^\nu - \nb^\nu h_\mu^\mu\right) 
- \frac{\kappa}{2} \nb_\mu \nb^\mu h_\nu^\nu + \cO(\kappa^2).
\ee
We consider the Einstein equation with a cosmological constant $\Lambda=\frac{3}{\ell^2}$,
\be
R_{\mu\nu} - \frac12 g_{\mu\nu} R +\frac{3}{\ell^2} g_{\mu\nu}=T_{\mu\nu},
\ee
where we use the natural unit $8\pi G_N=1$. At the linearized order $\cO(\kappa)$, the Einstein equation is reduced to\footnote{The expression of the linearized Einstein equation seems different from \cite{Date:2015kma}. But they are indeed the same.}
\begin{multline}\label{eom}
E_{\mu\nu}\equiv\frac12\left[\nb_\tau \nb_\mu h ^\tau_\nu + \nb_\tau \nb_\nu h ^\tau_\mu - \nb^2 h_{\mu\nu} - \nb_\mu \nb_\nu h\right]\\
-\frac12 \barg_{\mu\nu}\left(\nb_\alpha \nb_\beta h^{\alpha\beta} - \nb^2 h\right) - \frac{3}{\ell^2} h_{\mu\nu} + \frac{3}{2\ell^2} \barg_{\mu\nu} h = T_{\mu\nu},
\end{multline}
where we apply the relations from the Einstein equation of the dS spacetime
\be
\bar R_{\mu\nu}=\frac{3}{\ell^2} \barg_{\mu\nu},\quad\quad \bar R = \frac{12}{\ell^2},
\ee
and $h=\barg^{\mu\nu}h_{\mu\nu}$. The linearized equation \eqref{eom} is invariant under the gauge transformation
\be
h_{\mu\nu}\rightarrow h_{\mu\nu} + \nb_\mu \xi_\nu + \nb_\nu \xi_\mu.
\ee
To verify the gauge invariance, the following relations of the dS spacetime are applied,
\be
(\nb_\mu \nb_\nu - \nb_\nu \nb_\mu) \xi_\alpha={\bar R_{\mu\nu\alpha}}^{\quad\,\,\,\beta }\xi_\beta,\quad\quad \bar R_{\mu\nu\alpha\beta} =\frac{1}{\ell^2} (\barg_{\mu\alpha}\barg_{\nu\beta} -\barg_{\mu\beta}\barg_{\nu\alpha}).
\ee

\subsection{Near horizon symmetries in NU gauge}

We will work in the adapted NU gauge for linearized theory \cite{Conde:2016rom} where the following conditions are imposed,
\be
\begin{aligned}
&h_{rr}=h_{rz}=h_{r\bz}=h_{ru}=0\, ,\\
    &T_{rr}=T_{rz}=T_{r\bz}=T_{ru}=0.
\end{aligned}
\label{condition}
\ee
The radial components of the stress tensor $T_{r\mu}$ are set to zero to adapt to the gauge conditions of the perturbative metric, which can be done by introducing an auxiliary conserved symmetric 2-tensor \cite{Mao:2023dsy} as is detailed in Appendix \ref{stresstensor}.

The residual gauge transformation that preserves the conditions \eqref{condition} is generated by
\begin{align}
&\xi_u=Z(u,z,\bz)+r\p_u f(u,z,\bz)-F(r) f(u,z,\bz),\\
&\xi_r=-f(u,z,\bz),\\
&\xi_A=\Omega^2 Y_A(u,z,\bz) + \Omega D_A f(u,z,\bz) ,
\end{align}
where $D_A$ is the covariant derivative with respect to the unit sphere $\td s^2=\gamma_{AB}\td x^A \td x^B=\frac{4}{(1+z\bz)^2} \td z \td \bz$. The capital Latin indices are raised or lowered by the spherical metric $\gamma^{AB}$ and $\gamma_{AB}$.

We impose the following near horizon fall-off conditions 
\be
h_{z\bz}=\cO(r^2).
\ee
The absence of the leading order in $h_{z\bz}$ is to specify a traceless propagating mode. We impose a stronger traceless condition from a near horizon symmetry perspective. The fall-off conditions fix the independent symmetry parameter $Z$ that generates a translation along $r$ and select a $u$-independent near horizon supertranslation,
\be
Z=-\frac12 D_AD^A f(u,z,\bz)-\frac{\ell}{2} D_A Y^A,\quad\quad f(u,z,\bz)=T(z,\bz)+\frac12 \int \td u D_A Y^A. 
\ee
Now, $T$ characterizes the near horizon supertranslation and $Y^A$ is the near horizon superrotation. In vector form, it is given by
\begin{align}
&\xi^u=f , \\
&\xi^r=\frac12 D_A D^A f + \frac12 \Omega D_A Y^A , \\
&\xi^A= Y^A + \frac{1}{\Omega} D^A f.
\end{align}

\subsection{Near horizon solution space}

The organizations of Einstein equation in the Bondi gauge \cite{Bondi:1962px,Sachs:1962wk} and NU gauge for the case with a cosmological constant \cite{Compere:2019bua,Compere:2020lrt} are well known which greatly simplify the derivation of a solution space. Such configurations are inherited by linearized theory. Suppose that the metric components are given in series expansion near the horizon as
\begin{align}
&h_{AB}=N_{AB} + \sum_{i=1} N^{(i)}_{AB} r^i,\\
&h_{uA}=U_A+\Xi_A r + \sum_{i=2} U^{(i)}_A r^i,\\
&h_{uu}=V + \Theta r + \sum_{i=2} V^{(i)} r^i.
\end{align}
The first class of equations of motion is the radial part which completely determines the $r$-dependence of the trace of $h_{AB}$, $h_{uA}$, and $h_{uu}$ up to integration constants. In particular, $E_{rr}=0$ leads to
\be
\p_r^2\left(\frac{h_{z\bz}}{\ell-r}\right)=0.
\ee
The near horizon conditions $h_{z\bz}=\cO(r^2)$ yield $h_{z\bz}=0$. Then, $E_{rA}=0$ fix all the coefficients $U^{(i)}_A$ for $i\geq2$. The two leading orders in $g_{uA}$ are integration constants. Since the near horizon charges only involve the integration constants, we will not list the expressions for the higher orders. Next, $E_{ur}=0$ determines the coefficients $V^{(i)}$ for $i\geq2$, and 
\be
\Theta=\frac{1}{\ell} V + \frac{1}{2\ell^3}D_A D_B N^{AB} + \frac{1}{\ell^2}D_A U^A - \frac{1}{2\ell}D_A \Xi^A,
\ee
where the leading order $V$ is an integration constant. 

The second class of equations of motion is the standard equation which includes $E_{zz}=T_{zz}$ and $E_{\bz\bz}=T_{\bz\bz}$. They determine the time evolution of the traceless part of $N^{(i)}_{AB}$ for $i\geq1$. The leading order $N_{AB}$ is completely free, which we refer to as a news tensor in the near horizon analysis.
Once the first two classes of equations of motion are satisfied, the equation $E_{z\bz}=T_{z\bz}$ is fulfilled automatically. The other three equations are supplementary. The Bianchi identity at the linearized order guarantees that the supplementary equations are satisfied except for their leading orders, which yield the time evolution of the integration constants, 
\begin{multline}\label{A}
\ell^2 \p_u \Xi_A + 2\ell \p_u U_A  - 2 U_A - \ell^2 D_A \Theta\\
+D_B D_A U^B - D_B D^B U_A + \p_u D^B N_{AB}=2\ell^2 T_{uA}
\end{multline}
and
\be\label{u}
\ell^2\p_u V + D_A U^A - \frac{\ell}{2} D_A D^A V + \ell \p_u D_A U^A = \ell^3 T_{uu}.
\ee
It is important to point out that $U_A$ is also free. We will show that this extra freedom is very important for deriving a soft graviton theorem from the near horizon symmetry.

\subsection{Near horizon supertranslation charge}

For the (linearized) Einstein gravity, the surface charge associated to the near horizon symmetry is defined as \cite{Iyer:1994ys,Barnich:2001jy}
\begin{equation}
    {\mathcal{Q}}_\xi =\int_B\,\sqrt{-g} \Big( h^{\lambda [ u} \nabla _{\lambda} \xi^{r]} - \xi^{\lambda} \nabla^{[u} {h^{r]}}_{\lambda} - \frac{1}{2} h \nabla ^{[u} \xi^{r]} + \xi^{[u} \nabla _{\lambda} h^{r] \lambda} - \xi^{[u} \nabla^{r]}h \Big)\td z \td \bz,
\end{equation}
where we choose the bifurcation 2-sphere $B$ to evaluate the charge. In this work, we are mainly interested in the soft theorem associated to the near horizon supertranslation. Inserting the near horizon solution and the near horizon supertranslation, one can obtain
\be
{\mathcal{Q}}_T=\int_B\, \ga T \ell \left( V - \frac12 D_A \Xi^A \right).
\ee
The supertranslation charge can be evaluated on the whole horizon applying the relations in \eqref{A} and \eqref{u},
\begin{equation}
    \begin{aligned}
        Q_T=&\frac{1}{2 \ell}\int_{\mathcal{H}^-} \ga T\partial_u\left(D^AD^B N_{AB}\right)\td^2 z\td u\\
        &+ \ell \int_{\mathcal{H}^-} \ga T\left(\ell T_{uu}^0 - D^AT_{uA}^{0}\right)\td^2 z\td u\\
        &+\int_{\mathcal{H}^-}\ga T D^A\left(\frac12  D_A V - \frac{\ell}{2}  D_A \Theta - \frac{2}{\ell}  U_A \right)\td^2 z\td u .
\end{aligned}
\end{equation} 
The first line of this charge expression has the desired forms as a soft part of the charge \cite{Strominger:2013jfa,He:2014laa} for computing a soft theorem. The second line is in the form of a hard piece. We will turn off the third line by hand using the extra freedom from $U_A$ following the treatment in \cite{Mao:2023dsy}, for which we set
\be\label{extra}
\frac12 D_A V - \frac{\ell}{2}  D_A \Theta - \frac{2}{\ell} U_A =0.
\ee
The transformation laws of the supertranslation of the solutions, 
\begin{align}
&\delta_T U_A=-\frac12 D_A D_B D^B T,\\
&\delta_T \Xi_A=-\frac{2}{\ell} D_A T,\\
&\delta_T V=-\frac{1}{\ell} D_B D^B T,\\
&\delta_T \Theta = \frac{1}{\ell^2} D_B D^B T,\\
&\delta_T N_{AB}=2\ell( D_A D_B T - \frac12  \gamma_{AB}D^2 T ),
\end{align}
guarantee that the extra condition in \eqref{extra} is preserved by a supertranslation transformation.

The combination on the left hand side of \eqref{extra} was first noticed in \cite{Mao:2023dsy} for deriving a soft graviton theorem from the near horizon analysis in Schwarzschild spacetime. We believe this can be a universal structure in the near horizon analysis. The parameters $V$ and $\Theta$ in \eqref{extra} represent a non-propagating degree of freedom from the curvature of the spacetime. At null infinity, they do not contribute to the horizon charge. There is a natural split of the propagating degree of freedom and the non-propagating degree of freedom. But in the near horizon region, the degrees of freedom are mixed. Luckily, there is the extra degree of freedom from $U_A$ which could cancel those terms. The remarkable thing is that the special choice of $U_A$ is supertranslation invariant. Note that in the null infinity case, one integration constant is turned off by trivial diffeomorphism \cite{Bondi:1962px,Sachs:1962wk,Barnich:2010eb} in the metric component $g_{uA}$ which corresponds to the special configuration in \eqref{extra} in the near horizon case.  

Finally, we split the near horizon supertranslation charge into
\begin{align}
    Q_T=Q^S+Q^H ,
\end{align}
where
\begin{align}
   Q^S&=\frac{1}{2\ell}\int_{\mathcal{H}^-} \ga T\partial_u\left(D^AD^BN_{AB}\right)\td^2 z\td u, \\
   Q^H&=\ell\int_{\mathcal{H}^-} \ga T\left(\ell T_{uu}^0 - D^A T_{uA}^{0}\right)\td^2 z\td u,\label{hard}
\end{align}
are the soft and hard charges, respectively. In the null infinity analysis \cite{Strominger:2013jfa,He:2014laa}, it is normally assumed that the long-range magnetic mass aspect vanishes. Adapted to the near horizon case, this assumption is equivalent to imposing\footnote{Actually, it is not clear to us if this condition at the horizon is relevant to the magnetic mass aspect, because the near horizon charges are better appreciated from a thermodynamic perspective than the usual mass and angular momentum perspectives \cite{Adami:2021nnf,Adami:2021kvx}. The condition in \eqref{magnetic} is a direct translation from its null infinity counterpart. Of course, there is freedom to impose this condition in the near horizon case, i.e., it preserves the near horizon supertranslation.}
\begin{align}\label{magnetic}
    \left[D_zD_z N_{\bz\bz}-D_{\bz} D_{\bz} N_{zz}\right]_{\mathcal{H}^{-}_{\pm}}=0.
\end{align}
Then we can rewrite the soft charge as 
\begin{align}\label{soft supertranslation charge}
    Q^{S}=-\frac{1}{\ell}\int_{\mathcal{H}^-}  \partial_{\bz}T\partial_u\left(\frac{\partial_{\bz} N_{zz}}{\ga}\right)\td^2 z\td u.
\end{align}
For the hard part of the charge, it consists only the contribution from the stress tensor of the coupled matter fields. We have introduced an auxiliary field to modify the stress tensor which is equivalent to change the way of coupling the matter fields to gravity. As was shown in \cite{Mao:2023dsy}, the soft theorem derived from the Ward identity of the near horizon supertranslation charge will be dependent on the way of the matter fields couplings. The choice in \cite{Mao:2023dsy} is subject to a similar expression of the soft factor as the null infinity case. Here, we will follow the same treatment to further use the freedom in the auxiliary fields to turn off $T_{uA}^{0}$. The details of the modification are presented in Appendix \ref{stresstensor}. Then, the hard charge becomes
\be
Q^H=\ell^2\int_{\mathcal{H}^-} \ga T\, T_{uu}^0 \td^2 z\td u\, .
\ee


\section{Near horizon soft graviton theorem}
\label{soft}

In this section, we will demonstrate that the soft charge creates a low-energy soft graviton in near horizon states, and the action of the hard charge in those states leads to a soft factor, which mimics the scenario in flat spacetime at null infinity.

\subsection{Graviton modes in dS spacetime}

The mode expansion of the perturbative fields is the crucial ingredient for deriving a soft theorem from asymptotic symmetry, see, e.g. \cite{He:2014laa,Strominger:2017zoo}. It is very convenient to write down the mode expansion of the free field operator in the isotropic coordinates $(t,x_1,x_2,x_3)$ in curved spacetime \cite{Cheng:2022xgm,Cheng:2022xyr,Mao:2023rca,Mao:2023dsy}. The line element of the dS spacetime in the isotropic coordinates is
\be
\mathrm{d} s^{2}=-\left(\frac{\ell ^{2} -\rho ^{2}}{\ell ^{2} +\rho ^{2}}\right)^{2}\mathrm{d} t^{2} +\mathrm{\left(\frac{2\ell ^{2}}{\ell ^{2} +\rho ^{2}}\right)^{2} d}\vec{x} \cdot \mathrm{d}\vec{x}
,
\ee
where $\rho$ is related to the radial coordinate by
\be
r=\ell\frac{(\ell-\rho)^2}{\ell^2+\rho^2}.
\ee
The isotropic coordinates are connected to retarded coordinates $(u,r,z,\bz)$ by
\be
t =u+\frac{1}{2} \ell \log\left(\frac{r}{2\ell -r}\right),\quad x_1=\rho \frac{z+\bar{z}}{z\bar{z}+1}, \quad x_2=\rho\frac{z-\bar{z}}{i(z\bar{z}+1)},\quad x_3=\rho\frac{1-z\bar{z}}{z\bar{z}+1}.
\ee

In dS spacetime, $\frac{\partial}{\p t}$ is a timelike Killing vector which defines a positive energy of a particle as $\omega=-p_0$. One can write the dispersion relation for massless particles in the isotropic coordinates as 
\begin{align}
    -\left(\frac{\rho ^{2} +\ell ^{2}}{\rho ^{2} -\ell ^{2}}\right)^{2} \omega ^{2} +\left(\mathrm{\frac{\ell ^{2} +\rho ^{2}}{2\ell ^{2}}}\right)^{2}|\vec{p}|^{2}=0 .
\end{align}
Using the covariant measure of one-particle phase space 
\begin{align}
    \int\frac{\mathrm{d}\omega\mathrm{d}^3\vec{p}}{(2\pi)^3}\delta(p^\mu p_\mu)\theta(\omega)=\left(\frac{\ell ^{2} -\rho ^{2}}{\ell ^{2} +\rho ^{2}}\right)^{2}\int\frac{\mathrm{d}^3\vec{p}}{(2\pi)^3 2\omega},
\end{align}
a free massless scalar field $\phi(x)$ can be written in mode expansion as 
\begin{align}
    \phi(x^\mu)=\left(\frac{\ell ^{2} -\rho ^{2}}{\ell ^{2} +\rho ^{2}}\right)^{2}\int\frac{\mathrm{d}^3\vec{p}}{(2\pi)^3 2\omega}\left[a(\vec{p})e^{ip\cdot x}+a(\vec{p})^{\dagger}e^{-ip\cdot x}\right].
\end{align}
The extra factor $\left(\frac{\ell ^{2} -\rho ^{2}}{\ell ^{2} +\rho ^{2}}\right)^{2}$ is precisely from the dispersion relation. Here we consider only the correction from the dispersion relation of the null momentum, a full mode expansion of scalar field in dS spacetime can be performed perturbatively, see, e.g., in \cite{Bhatkar:2022qhz}. The mode expansion can be extended to a free graviton field simply by inserting the polarization tensor, 
\begin{align}
h_{\mu\nu}=\left(\frac{\ell ^{2} -\rho ^{2}}{\ell ^{2} +\rho ^{2}}\right)^{2}\sum_{\alpha=\pm}\int\frac{\td^3 p}{(2\pi)^3}\frac{1}{2\omega}\left[\epsilon^{\alpha*}_{\mu\nu }a_{\alpha}(\vec{p})e^{i\Vec{p}\cdot\Vec{x}}+\epsilon^{\alpha}_{\mu\nu }a^{\dagger}_{\alpha}(\vec{p})e^{-i\Vec{p}\cdot\Vec{x}}\right] ,
\end{align}
where $\epsilon_{\mu\nu}^\alpha$ is a product of the pair polarization vectors  $\epsilon_{\mu\nu}^\alpha=\epsilon^\alpha_\mu\epsilon^\alpha_\nu$.

One can parametrize the null momentum as
\begin{align}\label{parametrization}
    p_{\mu }=\frac{\omega }{1+z \bz}\frac{2\ell ^{2}}{\rho ^{2}-\ell ^{2}}\left(\frac{\ell ^{2} - \rho ^{2}}{2\ell ^{2}}( 1+z \bz) ,( z +\bz) ,-i( z -\bz) ,( 1-z \bz)\right) ,
\end{align}
and similarly the polarization vectors as 
\begin{align}
\epsilon ^{+\mu } & =\frac{1}{\sqrt{2}}\frac{\ell ^{2} +\rho ^{2}}{2\ell ^{2}}\left(\frac{2\ell ^{2}}{\rho ^{2}-\ell ^{2}}\bz ,1,-i,-\bz \right) ,\\
\epsilon ^{-\mu } & =\frac{1}{\sqrt{2}}\frac{\ell ^{2} +\rho ^{2}}{2\ell ^{2}}\left(\frac{2\ell ^{2}}{\rho ^{2}-\ell ^{2}} z ,1,i,-z \right) ,
\end{align}
which satisfies 
\begin{align}    \epsilon^{\alpha\mu}\epsilon^{\beta*}_{\mu}=\delta^{\alpha\beta}, \qquad p_\mu \epsilon^{\alpha\mu}=0 , \qquad \epsilon^{\alpha}_{r}=0.
\end{align}
Projecting the polarization vectors to the sphere, we obtain 
\be
\epsilon _{\bar{z}}^{+}  =\frac{2\sqrt{2} \ell ^{2} \rho }{( 1+z \bz)\left( \ell ^{2} +\rho ^{2}\right)} , \quad\quad \epsilon _{z}^{-}  =\frac{2\sqrt{2} \ell ^{2} \rho }{( 1+z \bz)\left( \ell ^{2} +\rho ^{2}\right)}.
\ee
Eventually, the near horizon field is related to the plane wave modes by
\begin{multline}\label{mode expansion}
    N_{zz}=\frac{4 \ga }{\pi ^{2}}\frac{\ell ^{8}(R+\ell)}{\left[\ell^2+\left(R+\ell\right)^2\right]^4} \int \mathrm{d} \omega  \left(\frac{r}{2\ell -r}\right)^{-i\ell \omega /2}\\
\times\sin\left[\frac{2\ell^2(R+\ell)\omega}{R(R+2\ell)}\right]  e^{\mathrm{-i\omega } u}a_{+}(\omega\hat{x}) + c.c. ,
\end{multline}
where $R=\rho-\ell$ and $R=0$ at the horizon. The integration of the three momentum $\vec{p}$ in the mode expansion in dS spacetime can be found in \cite{Mao:2023rca}. Literally, the near horizon field is obtained from the near horizon limit $R\rightarrow0$ of the mode expansion which is however singular. One can introduce a near horizon regularization $R\rightarrow R+i\mathcal{R}$ to deal with the divergence \cite{Cheng:2022xgm,Cheng:2022xyr,Mao:2023rca,Mao:2023dsy}. Nevertheless, we will keep the radial parameter for a moment and take the near horizon limit $R\rightarrow0$ at the last step. This is the reason we introduce a new radial parameter $R$ instead of simply using the previous radial parameter $r$.

\subsection{Soft graviton theorem in coordinate space}\label{soft theorem coordinates}

A soft graviton theorem can be derived from the Ward identity of supertranslation,
\begin{align}\label{ward identity}
    \left\langle {\rm out}\right|[Q_T,\mathcal{S}]\left|{\rm in}\right\rangle=0\Rightarrow Q^S\left|{\rm in}\right\rangle=-Q^H\left|{\rm in}\right\rangle ,
\end{align}
in the flat spacetime case \cite{He:2014laa,Strominger:2017zoo}. In the present analysis, we have omitted the out-part on $\mathcal{H}^+$, which can be easily restored from the $\mathcal{CPT}$ invariance. We choose $T=\frac{1}{z-w}$ for the supertranslation parameter, and thus 
\be 
\partial_{\bar{z}}T(z,\bar{z})=2\pi\delta^2(z-w).
\ee
With this choice, the soft charge \eqref{soft supertranslation charge} is reduced to 
\begin{equation}
Q^S=-\frac{4\pi^2i}{\ell}\lim_{\omega\rightarrow 0^+}\left[\omega\frac{\partial_{\bz}\Tilde{N}_{zz}}{\ga}\right] ,
\end{equation}
where $\Tilde{N}_{zz}$ is defined by a Fourier relation 
\be
N_{zz}=\int_{-\infty}^{+\infty}\td u e^{i\omega u} \Tilde{N}_{zz},
\ee
which implies that
\begin{equation}
\int_{-\infty}^{\infty}\td u\,\partial_u {N}_{zz}(u)=2\pi i\lim_{\omega\to0}\left[\omega\tilde{{N}}_{zz}(\omega)\right].
\end{equation}
Inserting the mode expansion \eqref{mode expansion}, one can obtain 
\begin{align}
     Q^S =-i\frac{\ell^2}{R}\frac{\partial_{\bz}\ga}{\ga}\lim_{\omega\rightarrow0^+}\left[\omega^2a_{+}+\omega^2a_{-}^\dagger\right] .
\end{align}
Hence, the soft charge $Q^S$ acts on the in-state as 
\begin{align}\label{soft charge act}
    Q^S\left|{\rm in}\right\rangle=-i\frac{\ell^2}{R}\frac{\partial_{\bz}\ga}{\ga}\lim_{\omega\rightarrow 0^+}\omega^2a_{-}^\dagger\left|{\rm in}\right\rangle.
\end{align}
The soft charge creates a low-energy soft graviton in the near horizon region.

For a massless scalar field coupled case as described in Appendix \ref{stresstensor}, one has the commutation relation \cite{Mao:2023rca},
\be
[\partial_u\bar{\phi}(u,z,\bz),\phi(g,z,\bz)]=-\frac{i}{\ell^2\ga}\delta(u-g)\delta^2(z-w) ,
\ee
which yields the action of the hard charge as
\be
[Q^H,\phi]=-i T\partial_u\phi .
\ee
The above relation and the special choice of the supertranslation parameter $T=\frac{1}{z-w}$ determine the action of the hard charge $Q^H$ on the in-state as  
\begin{align}\label{hard charge act}
   Q^H \left|{\rm in}\right\rangle= \sum_{k=1}^{n}\frac{E_k}{(z-z_k)}\left|{\rm in}\right\rangle\, .
\end{align}
Finally, we can obtain a soft graviton theorem in coordinate space by inserting \eqref{soft charge act} and \eqref{hard charge act} into the Ward identity \eqref{ward identity},
\be
\label{soft theorem coor}
\partial_{\bz}\left[\ga\lim_{\omega\rightarrow 0^+}a^\dagger_{-}\left|{\rm in}\right\rangle\right]=\lim_{\omega\rightarrow 0^+}\left[\frac{\pi}{i}\frac{ R}{\ell^2\omega}\ga\sum_{k=1}^n\frac{E_k}{\omega(z-z_k)}\left|{\rm in}\right\rangle\right] .
\ee

\subsection{Soft graviton theorem in momentum space}

In this subsection, we will use the null parametrization in \eqref{parametrization} to rewrite the soft theorem in momentum space. For each hard particle, their momenta can be parametrized as 
\begin{align}
    q_{k\mu }^{\mathrm{in}}=\frac{E_k^{\mathrm{in}}}{1+z_k^\mathrm{in}\bar{z }_k^\mathrm{in}}\frac{2\ell ^{2}}{\rho ^{2}-\ell ^{2}}\left(\frac{\ell ^{2} -\rho ^{2}}{2\ell ^{2}}( 1+z_k^\mathrm{in} \bar{z }_k^\mathrm{in}) ,( z_k^\mathrm{in} +\bar{z }_k^\mathrm{in}) ,-i( z_k^\mathrm{in} -\bar{z }_k^\mathrm{in}) ,( 1-z_k^\mathrm{in} \bar{z }_k^\mathrm{in})\right) .
\end{align}
One can easily verify that
\begin{equation}\label{soft factor momentum}
    \begin{aligned}
    \partial_{\bz}\left(\ga\sum_{k=1}^n\frac{[q_k\cdot \epsilon^+(p)]^2}{p\cdot q_k}\right)&=-\ga\sum_{k=1}^n\frac{E_k}{\omega(z-z_k)}\left(  1+\frac{\bz_k(z-z_k)}{1+z_k\bz_k}\right)\\
    &=-\ga\sum_{k=1}^n\frac{E_k}{\omega(z-z_k)} ,
\end{aligned}
\end{equation}
where we have used the conversation of a combination of two components of the total momentum,
\begin{align}
   \sum_{k=1}^n \frac{E_k\bz_k}{1+z_k\bz_k}\propto\sum_{k=1}^n \left(q_{k1}-iq_{k2}\right)=0 .
\end{align}
Applying the relation \eqref{soft factor momentum}, we arrive at a soft graviton theorem in the momentum space as
\begin{equation}\label{softmomentum}
    \begin{aligned}
    \lim_{\omega\rightarrow 0^+}&\langle {\rm out}|a_{+}\mathcal{S}-\mathcal{S}a^\dagger_{-}\left|{\rm in}\right\rangle\\
    &=\lim_{\omega\rightarrow 0^+}i\frac{\pi R}{\omega \ell^2}\left[\sum_{l=1}^m\frac{(q_l^{\rm out}\cdot \epsilon)^2}{p\cdot q_l^{\rm out}}-\sum_{k=1}^n\frac{(q_k^{\rm in}\cdot \epsilon)^2}{p\cdot q_k^{\rm in}}\right]\langle{\rm out}|\mathcal{S}\left|{\rm in}\right\rangle,
\end{aligned}
\end{equation}
where the derivative $\p_{\bz}$ was dropped from both sides. One can see that the soft factor on the right hand side of \eqref{softmomentum} contains the flat spacetime soft factor and a prefactor $i\frac{\pi R}{\omega \ell^2}$. A flat limit can be taken by setting $\frac{\pi i R}{\omega \ell^2}=1$. Physically, this limit can be understood as the fact that the soft limit and the flat limit of the cosmological constant $\Lambda$ are at the same order tending to zero. Note that $iR$ is from the near horizon regularization. The orders of the near horizon limit and the flat limit for the parameter $R=\rho-\ell$ do not commute. The present work is based on the near horizon analysis. So we first take the near horizon limit. However this will somehow prevent the flat limit since the flat limit means the vanish of the cosmological horizon. Thus, the horizon regularization $R\rightarrow R-i\mathcal{R}$ also ensures a flat limit for the soft theorem in the momentum space. Now the regularization involves a minus sign as $\rho\leq\ell$. Then, the flat limit condition becomes $\Lambda=\frac{\omega}{ \pi  \mathcal{R} }$.


\section{Conclusion and discussion}

In this paper, we study the linearized gravity theory in the near horizon region of the dS spacetime. The near horizon symmetry, near horizon solution space, and near horizon supertranslation charge are obtained. A soft graviton theorem in dS spacetime is derived from the Ward identity of the near horizon supertranslation. The soft theorem has a similar structure as the flat spacetime soft theorem which relies on a fine tuning structure of the near horizon fall-off conditions. The remarkable feature of this fine tuning structure resides in its supertranslation invariance. The soft theorem in dS spacetime has a well defined flat limit, which recovers a flat spacetime soft theorem.

To close this paper, we would like to comment on some subtleties and future directions. In the near horizon analysis of the black hole case, the soft limit should involve also the mass parameter of the black hole rather than simply comparing the soft and hard external particles \cite{Gaddam:2020mwe,Gaddam:2020rxb}. Here, our derivation simply involves the soft and hard external particles. We think this is a less subtle issue in dS spacetime. The reasoning is that one can consider a perturbative expansion in the inverse of the curvature length scale $\frac{1}{\ell}$ \cite{Bhatkar:2022qhz}. The leading contribution should be the flat spacetime result. This is achieved from the flat limit of the soft theorem in momentum space \eqref{softmomentum}. It is meaningful to point out that our derivation is from the near horizon analysis in dS spacetime. The final result is indeed the leading term of the dS spacetime as it should be. In some sense, we have not considered the $\frac{1}{\ell}$ corrections from near horizon analysis, which is definitely an interesting direction for future investigation. In particular, holography provides a powerful alternative approach to investigate the soft modes and flat limit in the spacetime with a cosmological constant \cite{Duary:2022afn}. For other future directions, the obvious one is to consider the full interacting Einstein gravity theory as the generic near horizon symmetry has already been derived in \cite{Adami:2021nnf}. While the relevant investigations are mainly dealing with soft limit of massless particles, in Einstein theory, general covariance prevents a local definition of energy. Hence, soft limit is a very subtle point.\footnote{The situation at infinity is very different. One normally defines the in or out states for massless interaction at the null infinity. Since any asymptotically flat spacetime has the same structure of null infinity, it is reasonable to consider the energy with respect to Minkowski spacetime as the energy of the massless particle near null infinity.} Nevertheless, one can extend the present study to include self-interaction between (near horizon) gravitons. The stress tensor will include self-interaction terms which is somewhat similar to the case of Yang-Mills theory \cite{Cheng:2022xgm,Mao:2023rca}. Technically, we expect the near horizon symmetry and charge analysis would be just a special case of \cite{Adami:2021nnf}, i.e., the null boundary should be fixed as the dS horizon. There are also some other interesting future directions such as the subleading soft theorem in the low-energy expansion, i.e., the dS analog of the investigations in \cite{Lysov:2014csa,Campiglia:2014yka,Campiglia:2016jdj,Conde:2016rom,Campiglia:2016efb}. A more challenging point is the dual interpretation from the point of view of celestial holography \cite{Pasterski:2016qvg,Pasterski:2023ikd}. In particular, there are some recent studies on the deformations of the soft theorem \cite{He:2022zcf,He:2023lvk,He:2024yzx}, which may also be extended to the case of soft theorem in curved spacetime.


\section*{Acknowledgments}
The authors thank Geoffrey Comp\`{e}re for useful correspondence on the linearized Einstein equation. The authors would like to thank Kai-Yu Zhang for useful discussions and collaborations in relevant research topics. This work is supported in part by the National Natural Science Foundation of China (NSFC) under Grants No. 11935009 and No. 11905156.

\appendix
\section{Modification of stress tensor}
\label{stresstensor}

As was shown in \cite{Mao:2023dsy}, one can modify the stress tensor by adding a divergence-free symmetric rank two tensor. We will modify the stress tensor to satisfy the gauge condition in \eqref{condition}. In this work, we will limit ourselves to a massless complex scalar field which is originally minimally coupled to gravity. Thus, the stress tensor is
\begin{align}
    \Tilde{T}_{\mu\nu}\equiv \frac{1}{\sqrt{-g}}\frac{\delta(\sqrt{-g}\mathcal{L}_{M})}{\delta g_{\mu\nu}}=\frac{1}{2}\left(\partial_\mu\Phi\partial_\nu\bar{\Phi}+\partial_\nu\Phi\partial_\mu\bar{\Phi}\right)-\frac{1}{2}g_{\mu\nu}\nb^\rho\Phi\nb_\rho\bar{\Phi}\, ,
\end{align}
where   $\mathcal{L}_M=\nb_\mu\Phi\nb^\mu\bar{\Phi}\,$. We assume that the scalar fields are given in the form of near horizon expansions as  
\begin{align}
    \Phi=\phi+\sum_{n=1}^{\infty}r^n \Phi^{(n)}\, ,\quad 
\bar{\Phi}=\bar{\phi}+\sum_{n=1}^{\infty}r^n \bar{\Phi}^{(n)}\, .
\end{align}
We construct a modified stress tensor by $T_{\mu\nu}=\Tilde{T}_{\mu\nu}-\Upsilon_{\mu\nu}$, where $T_{\mu\nu}$ satisfies the gauge conditions in \eqref{condition} and $\Upsilon_{\mu\nu}$ is an arbitrary tensor. Then we will use the divergence-free conditions $\nb^\mu \Upsilon_{\mu\nu}=0$ to fix it.
From $\nb^\mu\Upsilon_{\mu r}=0$, one obtains
\be
    \gamma^{AB}\Upsilon_{AB}=\Omega^3 \partial_u \Tilde{T}_{rr} - \Omega D^A\Tilde{T}_{rA}
+  \Omega^3 \partial_r\Tilde{T}_{ur} - 2 \Omega^2 \tilde{T}_{ur} + 2 F \Omega^2 \tilde{T}_{rr}-\Omega^3 F \p_r \tilde{T}_{rr} - 3 \frac{\Omega^4}{\ell^2 } \tilde{T}_{rr},
\ee
where we have used the relation $\Upsilon_{r\mu}=\Tilde{T}_{r\mu}$ to fulfill the gauge conditions for $T_{\mu\nu}$. Thus the trace part of $\Upsilon_{\mu\nu}$ is completely fixed by $\Tilde{T}_{\mu\nu}$.

The transverse equations $\nb^\mu\Upsilon_{\mu A}=0$ yield
\be
\partial_r\Upsilon_{uA}=F\p_r \Tilde{T}_{rA} -\partial_u\Tilde{T}_{rA} + \frac{2}{\Omega}\Upsilon_{uA} + 
(\frac{2\Omega}{\ell^2}-\frac{2F}{\Omega} )\Tilde{T}_{rA} + \frac{1}{\Omega^2} D^B \Upsilon_{AB}.
\ee
Suppose that $\Upsilon_{uA}$ is given by the series 
\begin{align}
\Upsilon_{uA}=\Upsilon^0_{uA}+\sum_{n=1}^\infty r^n \Upsilon_{uA}^{(n)}\, .
\end{align}
All the order $\Upsilon_{uA}^{(n)}$ for $n\geq1$ is fixed from the above equation. While the leading order $\Upsilon^{0}_{uA}$ is free as an integration constant. We continue with the $u$ component of the divergence-free conditions, which gives
\be
\partial_r\Upsilon_{uu}=F \p_r \Tilde{T}_{ru} - \partial_u\Tilde{T}_{ru} + (\frac{2}{\ell^2}\Omega - \frac{2}{\Omega}F) \Tilde{T}_{ru} 
+ \frac{1}{\Omega^2} D^A\Upsilon_{uA} + \frac{2}{\Omega} \Upsilon_{uu}.
\ee
This equation controls $\Upsilon_{uu}$ up to an integration constant $\Upsilon^0_{uu}$ at the leading order in the near horizon expansion. Finally, we find that the traceless part of $\Upsilon_{AB}$ and the leading order of $\Upsilon_{uu}$, $\Upsilon_{uA}$ of the auxiliary tensor are free as initial data and the rest ingredients are fixed by the gauge conditions. The $uu, uA$ components of the modified stress tensor at the leading orders are
\begin{align}
T_{uu}^{0}&=\partial_u\phi\partial_u\bar{\phi}-\Upsilon_{uu}^0 ,\\
    T_{uA}^{0}&=\frac{1}{2}\left(\partial_u\phi\partial_A\bar{\phi}+\partial_A\phi\partial_u\bar{\phi}\right)-\Upsilon_{uA}^0.
\end{align}
The hard part of the near horizon supertranslation charge \eqref{hard} is indeed sensitive to the modification of the stress tensor. In this work, we follow the previous choice in \cite{Mao:2023dsy} to set $\Upsilon_{uu}^0=0$ and $T_{uA}^{0}=0$.

\providecommand{\href}[2]{#2}\begingroup\raggedright\endgroup


\begin{thebibliography}{10}

\bibitem{Hawking:2016msc}
S.~W. Hawking, M.~J. Perry, and A.~Strominger, ``{Soft Hair on Black Holes},''
  \href{http://dx.doi.org/10.1103/PhysRevLett.116.231301}{{\em Phys. Rev.
  Lett.} {\bfseries 116} no.~23, (2016) 231301},
  \href{http://arxiv.org/abs/1601.00921}{{\ttfamily arXiv:1601.00921
  [hep-th]}}.

\bibitem{Donnay:2015abr}
L.~Donnay, G.~Giribet, H.~A. Gonzalez, and M.~Pino, ``{Supertranslations and
  Superrotations at the Black Hole Horizon},''
  \href{http://dx.doi.org/10.1103/PhysRevLett.116.091101}{{\em Phys. Rev.
  Lett.} {\bfseries 116} no.~9, (2016) 091101},
  \href{http://arxiv.org/abs/1511.08687}{{\ttfamily arXiv:1511.08687
  [hep-th]}}.

\bibitem{Averin:2016ybl}
A.~Averin, G.~Dvali, C.~Gomez, and D.~Lust, ``{Gravitational Black Hole Hair
  from Event Horizon Supertranslations},''
  \href{http://dx.doi.org/10.1007/JHEP06(2016)088}{{\em JHEP} {\bfseries 06}
  (2016) 088}, \href{http://arxiv.org/abs/1601.03725}{{\ttfamily
  arXiv:1601.03725 [hep-th]}}.

\bibitem{Afshar:2016wfy}
H.~Afshar, S.~Detournay, D.~Grumiller, W.~Merbis, A.~Perez, D.~Tempo, and
  R.~Troncoso, ``{Soft Heisenberg hair on black holes in three dimensions},''
  \href{http://dx.doi.org/10.1103/PhysRevD.93.101503}{{\em Phys. Rev. D}
  {\bfseries 93} no.~10, (2016) 101503},
  \href{http://arxiv.org/abs/1603.04824}{{\ttfamily arXiv:1603.04824
  [hep-th]}}.

\bibitem{Setare:2016jba}
M.~R. Setare and H.~Adami, ``{Near Horizon Symmetries of the Non-Extremal Black
  Hole Solutions of Generalized Minimal Massive Gravity},''
  \href{http://dx.doi.org/10.1016/j.physletb.2016.07.022}{{\em Phys. Lett. B}
  {\bfseries 760} (2016) 411--416},
  \href{http://arxiv.org/abs/1606.02273}{{\ttfamily arXiv:1606.02273
  [hep-th]}}.

\bibitem{Mao:2016pwq}
P.~Mao, X.~Wu, and H.~Zhang, ``{Soft hairs on isolated horizon implanted by
  electromagnetic fields},''
  \href{http://dx.doi.org/10.1088/1361-6382/aa59da}{{\em Class. Quant. Grav.}
  {\bfseries 34} no.~5, (2017) 055003},
  \href{http://arxiv.org/abs/1606.03226}{{\ttfamily arXiv:1606.03226
  [hep-th]}}.

\bibitem{Setare:2016vhy}
M.~R. Setare and H.~Adami, ``{The Heisenberg algebra as near horizon symmetry
  of the black flower solutions of Chern\textendash{}Simons-like theories of
  gravity},'' \href{http://dx.doi.org/10.1016/j.nuclphysb.2016.11.011}{{\em
  Nucl. Phys. B} {\bfseries 914} (2017) 220--233},
  \href{http://arxiv.org/abs/1606.05260}{{\ttfamily arXiv:1606.05260
  [hep-th]}}.

\bibitem{Afshar:2016uax}
H.~Afshar, D.~Grumiller, and M.~M. Sheikh-Jabbari, ``{Near horizon soft hair as
  microstates of three dimensional black holes},''
  \href{http://dx.doi.org/10.1103/PhysRevD.96.084032}{{\em Phys. Rev. D}
  {\bfseries 96} no.~8, (2017) 084032},
  \href{http://arxiv.org/abs/1607.00009}{{\ttfamily arXiv:1607.00009
  [hep-th]}}.

\bibitem{Grumiller:2016kcp}
D.~Grumiller, A.~Perez, S.~Prohazka, D.~Tempo, and R.~Troncoso, ``{Higher Spin
  Black Holes with Soft Hair},''
  \href{http://dx.doi.org/10.1007/JHEP10(2016)119}{{\em JHEP} {\bfseries 10}
  (2016) 119}, \href{http://arxiv.org/abs/1607.05360}{{\ttfamily
  arXiv:1607.05360 [hep-th]}}.

\bibitem{Donnay:2016ejv}
L.~Donnay, G.~Giribet, H.~A. Gonz\'alez, and M.~Pino, ``{Extended Symmetries at
  the Black Hole Horizon},''
  \href{http://dx.doi.org/10.1007/JHEP09(2016)100}{{\em JHEP} {\bfseries 09}
  (2016) 100}, \href{http://arxiv.org/abs/1607.05703}{{\ttfamily
  arXiv:1607.05703 [hep-th]}}.

\bibitem{Setare:2016msj}
M.~R. Setare and H.~Adami, ``{BMS type symmetries at null-infinity and near
  horizon of non-extremal black holes},''
  \href{http://dx.doi.org/10.1140/epjc/s10052-016-4548-0}{{\em Eur. Phys. J. C}
  {\bfseries 76} no.~12, (2016) 687},
  \href{http://arxiv.org/abs/1609.05736}{{\ttfamily arXiv:1609.05736
  [hep-th]}}.

\bibitem{Sheikh-Jabbari:2016npa}
M.~M. Sheikh-Jabbari and H.~Yavartanoo, ``{Horizon Fluffs: Near Horizon Soft
  Hairs as Microstates of Generic AdS3 Black Holes},''
  \href{http://dx.doi.org/10.1103/PhysRevD.95.044007}{{\em Phys. Rev. D}
  {\bfseries 95} no.~4, (2017) 044007},
  \href{http://arxiv.org/abs/1608.01293}{{\ttfamily arXiv:1608.01293
  [hep-th]}}.

\bibitem{Cai:2016idg}
R.-G. Cai, S.-M. Ruan, and Y.-L. Zhang, ``{Horizon supertranslation and
  degenerate black hole solutions},''
  \href{http://dx.doi.org/10.1007/JHEP09(2016)163}{{\em JHEP} {\bfseries 09}
  (2016) 163}, \href{http://arxiv.org/abs/1609.01056}{{\ttfamily
  arXiv:1609.01056 [gr-qc]}}.

\bibitem{Hawking:2016sgy}
S.~W. Hawking, M.~J. Perry, and A.~Strominger, ``{Superrotation Charge and
  Supertranslation Hair on Black Holes},''
  \href{http://dx.doi.org/10.1007/JHEP05(2017)161}{{\em JHEP} {\bfseries 05}
  (2017) 161}, \href{http://arxiv.org/abs/1611.09175}{{\ttfamily
  arXiv:1611.09175 [hep-th]}}.

\bibitem{Afshar:2016kjj}
H.~Afshar, D.~Grumiller, W.~Merbis, A.~Perez, D.~Tempo, and R.~Troncoso,
  ``{Soft hairy horizons in three spacetime dimensions},''
  \href{http://dx.doi.org/10.1103/PhysRevD.95.106005}{{\em Phys. Rev. D}
  {\bfseries 95} no.~10, (2017) 106005},
  \href{http://arxiv.org/abs/1611.09783}{{\ttfamily arXiv:1611.09783
  [hep-th]}}.

\bibitem{Shi:2016jtn}
C.~Shi and J.~Mei, ``{Extended Symmetries at Black Hole Horizons in Generic
  Dimensions},'' \href{http://dx.doi.org/10.1103/PhysRevD.95.104053}{{\em Phys.
  Rev. D} {\bfseries 95} no.~10, (2017) 104053},
  \href{http://arxiv.org/abs/1611.09491}{{\ttfamily arXiv:1611.09491 [gr-qc]}}.

\bibitem{Eling:2016qvx}
C.~Eling, ``{Spontaneously Broken Asymptotic Symmetries and an Effective Action
  for Horizon Dynamics},''
  \href{http://dx.doi.org/10.1007/JHEP02(2017)052}{{\em JHEP} {\bfseries 02}
  (2017) 052}, \href{http://arxiv.org/abs/1611.10214}{{\ttfamily
  arXiv:1611.10214 [hep-th]}}.

\bibitem{Akhmedov:2017ftb}
E.~T. Akhmedov and M.~Godazgar, ``{Symmetries at the black hole horizon},''
  \href{http://dx.doi.org/10.1103/PhysRevD.96.104025}{{\em Phys. Rev. D}
  {\bfseries 96} no.~10, (2017) 104025},
  \href{http://arxiv.org/abs/1707.05517}{{\ttfamily arXiv:1707.05517
  [hep-th]}}.

\bibitem{Grumiller:2019tyl}
D.~Grumiller and W.~Merbis, ``{Near horizon dynamics of three dimensional black
  holes},'' \href{http://dx.doi.org/10.21468/SciPostPhys.8.1.010}{{\em SciPost
  Phys.} {\bfseries 8} no.~1, (2020) 010},
  \href{http://arxiv.org/abs/1906.10694}{{\ttfamily arXiv:1906.10694
  [hep-th]}}.

\bibitem{Grumiller:2019fmp}
D.~Grumiller, A.~P\'erez, M.~M. Sheikh-Jabbari, R.~Troncoso, and C.~Zwikel,
  ``{Spacetime structure near generic horizons and soft hair},''
  \href{http://dx.doi.org/10.1103/PhysRevLett.124.041601}{{\em Phys. Rev.
  Lett.} {\bfseries 124} no.~4, (2020) 041601},
  \href{http://arxiv.org/abs/1908.09833}{{\ttfamily arXiv:1908.09833
  [hep-th]}}.

\bibitem{Adami:2020amw}
H.~Adami, D.~Grumiller, S.~Sadeghian, M.~M. Sheikh-Jabbari, and C.~Zwikel,
  ``{T-Witts from the horizon},''
  \href{http://dx.doi.org/10.1007/JHEP04(2020)128}{{\em JHEP} {\bfseries 04}
  (2020) 128}, \href{http://arxiv.org/abs/2002.08346}{{\ttfamily
  arXiv:2002.08346 [hep-th]}}.

\bibitem{Grumiller:2020vvv}
D.~Grumiller, M.~M. Sheikh-Jabbari, and C.~Zwikel, ``{Horizons 2020},''
  \href{http://dx.doi.org/10.1142/S0218271820430063}{{\em Int. J. Mod. Phys. D}
  {\bfseries 29} no.~14, (2020) 2043006},
  \href{http://arxiv.org/abs/2005.06936}{{\ttfamily arXiv:2005.06936
  [hep-th]}}.

\bibitem{Donnay:2020yxw}
L.~Donnay, G.~Giribet, and J.~Oliva, ``{Horizon symmetries and hairy black
  holes in AdS},'' \href{http://dx.doi.org/10.1007/JHEP09(2020)120}{{\em JHEP}
  {\bfseries 09} (2020) 120}, \href{http://arxiv.org/abs/2007.08422}{{\ttfamily
  arXiv:2007.08422 [hep-th]}}.

\bibitem{Adami:2020ugu}
H.~Adami, M.~M. Sheikh-Jabbari, V.~Taghiloo, H.~Yavartanoo, and C.~Zwikel,
  ``{Symmetries at null boundaries: two and three dimensional gravity cases},''
  \href{http://dx.doi.org/10.1007/JHEP10(2020)107}{{\em JHEP} {\bfseries 10}
  (2020) 107}, \href{http://arxiv.org/abs/2007.12759}{{\ttfamily
  arXiv:2007.12759 [hep-th]}}.

\bibitem{Adami:2021sko}
H.~Adami, M.~M. Sheikh-Jabbari, V.~Taghiloo, H.~Yavartanoo, and C.~Zwikel,
  ``{Chiral Massive News: Null Boundary Symmetries in Topologically Massive
  Gravity},'' \href{http://dx.doi.org/10.1007/JHEP05(2021)261}{{\em JHEP}
  {\bfseries 05} (2021) 261}, \href{http://arxiv.org/abs/2104.03992}{{\ttfamily
  arXiv:2104.03992 [hep-th]}}.

\bibitem{Adami:2021nnf}
H.~Adami, D.~Grumiller, M.~M. Sheikh-Jabbari, V.~Taghiloo, H.~Yavartanoo, and
  C.~Zwikel, ``{Null boundary phase space: slicings, news \& memory},''
  \href{http://dx.doi.org/10.1007/JHEP11(2021)155}{{\em JHEP} {\bfseries 11}
  (2021) 155}, \href{http://arxiv.org/abs/2110.04218}{{\ttfamily
  arXiv:2110.04218 [hep-th]}}.

\bibitem{Adami:2021kvx}
H.~Adami, M.~M. Sheikh-Jabbari, V.~Taghiloo, and H.~Yavartanoo, ``{Null surface
  thermodynamics},'' \href{http://dx.doi.org/10.1103/PhysRevD.105.066004}{{\em
  Phys. Rev. D} {\bfseries 105} no.~6, (2022) 066004},
  \href{http://arxiv.org/abs/2110.04224}{{\ttfamily arXiv:2110.04224
  [hep-th]}}.

\bibitem{Liu:2022uox}
H.-S. Liu and P.~Mao, ``{Near horizon gravitational charges},''
  \href{http://dx.doi.org/10.1007/JHEP05(2022)123}{{\em JHEP} {\bfseries 05}
  (2022) 123}, \href{http://arxiv.org/abs/2201.10308}{{\ttfamily
  arXiv:2201.10308 [hep-th]}}.

\bibitem{Adami:2022ktn}
H.~Adami, P.~Mao, M.~M. Sheikh-Jabbari, V.~Taghiloo, and H.~Yavartanoo,
  ``{Symmetries at causal boundaries in 2D and 3D gravity},''
  \href{http://dx.doi.org/10.1007/JHEP05(2022)189}{{\em JHEP} {\bfseries 05}
  (2022) 189}, \href{http://arxiv.org/abs/2202.12129}{{\ttfamily
  arXiv:2202.12129 [hep-th]}}.

\bibitem{Taghiloo:2022hxc}
V.~Taghiloo, ``{Null surface thermodynamics in topologically massive
  gravity},'' \href{http://dx.doi.org/10.1140/epjc/s10052-023-11309-0}{{\em
  Eur. Phys. J. C} {\bfseries 83} no.~2, (2023) 182},
  \href{http://arxiv.org/abs/2205.10909}{{\ttfamily arXiv:2205.10909
  [hep-th]}}.

\bibitem{Mao:2022ldv}
P.~Mao and W.~Zhao, ``{Null boundary gravitational charges from local Lorentz
  symmetries},'' \href{http://dx.doi.org/10.1103/PhysRevD.107.044004}{{\em
  Phys. Rev. D} {\bfseries 107} no.~4, (2023) 044004},
  \href{http://arxiv.org/abs/2211.04736}{{\ttfamily arXiv:2211.04736
  [hep-th]}}.

\bibitem{Aggarwal:2023qwl}
A.~Aggarwal and N.~Gaddam, ``{All symmetries of near-horizon scattering},''
  \href{http://arxiv.org/abs/2309.05775}{{\ttfamily arXiv:2309.05775
  [hep-th]}}.

\bibitem{Ashtekar:2024mme}
A.~Ashtekar and S.~Speziale, ``{Horizons and null infinity: A fugue in four
  voices},'' \href{http://dx.doi.org/10.1103/PhysRevD.109.L061501}{{\em Phys.
  Rev. D} {\bfseries 109} no.~6, (2024) L061501},
  \href{http://arxiv.org/abs/2401.15618}{{\ttfamily arXiv:2401.15618 [gr-qc]}}.

\bibitem{Ashtekar:2024bpi}
A.~Ashtekar and S.~Speziale, ``{Null Infinity as a Weakly Isolated Horizon},''
  \href{http://arxiv.org/abs/2402.17977}{{\ttfamily arXiv:2402.17977
  [hep-th]}}.

\bibitem{Strominger:2017zoo}
A.~Strominger, {\em {Lectures on the Infrared Structure of Gravity and Gauge
  Theory}}.
\newblock Princeton University Press, Princeton, 2018.
\newblock \href{http://arxiv.org/abs/1703.05448}{{\ttfamily arXiv:1703.05448
  [hep-th]}}.

\bibitem{Donnay:2018ckb}
L.~Donnay, G.~Giribet, H.~A. Gonz\'alez, and A.~Puhm, ``{Black hole memory
  effect},'' \href{http://dx.doi.org/10.1103/PhysRevD.98.124016}{{\em Phys.
  Rev. D} {\bfseries 98} no.~12, (2018) 124016},
  \href{http://arxiv.org/abs/1809.07266}{{\ttfamily arXiv:1809.07266
  [hep-th]}}.

\bibitem{Rahman:2019bmk}
A.~A. Rahman and R.~M. Wald, ``{Black Hole Memory},''
  \href{http://dx.doi.org/10.1103/PhysRevD.101.124010}{{\em Phys. Rev. D}
  {\bfseries 101} no.~12, (2020) 124010},
  \href{http://arxiv.org/abs/1912.12806}{{\ttfamily arXiv:1912.12806 [gr-qc]}}.

\bibitem{Bhattacharjee:2020vfb}
S.~Bhattacharjee, S.~Kumar, and A.~Bhattacharyya, ``{Displacement memory effect
  near the horizon of black holes},''
  \href{http://dx.doi.org/10.1007/JHEP03(2021)134}{{\em JHEP} {\bfseries 03}
  (2021) 134}, \href{http://arxiv.org/abs/2010.16086}{{\ttfamily
  arXiv:2010.16086 [gr-qc]}}.

\bibitem{Cheng:2022xyr}
P.~Cheng and P.~Mao, ``{Soft theorems in curved spacetime},''
  \href{http://dx.doi.org/10.1103/PhysRevD.106.L081702}{{\em Phys. Rev. D}
  {\bfseries 106} no.~8, (2022) L081702},
  \href{http://arxiv.org/abs/2206.11564}{{\ttfamily arXiv:2206.11564
  [hep-th]}}.

\bibitem{Cheng:2022xgm}
P.~Cheng and P.~Mao, ``{Soft gluon theorems in curved spacetime},''
  \href{http://dx.doi.org/10.1103/PhysRevD.107.065010}{{\em Phys. Rev. D}
  {\bfseries 107} no.~6, (2023) 065010},
  \href{http://arxiv.org/abs/2211.00031}{{\ttfamily arXiv:2211.00031
  [hep-th]}}.

\bibitem{Mao:2023dsy}
P.~Mao, K.-Y. Zhang, and B.~Zhou, ``{Near horizon linearized gravity and soft
  theorem},'' \href{http://dx.doi.org/10.1103/PhysRevD.109.065022}{{\em Phys.
  Rev. D} {\bfseries 109} no.~6, (2024) 065022},
  \href{http://arxiv.org/abs/2311.03773}{{\ttfamily arXiv:2311.03773
  [hep-th]}}.

\bibitem{Mao:2023rca}
P.~Mao and K.-Y. Zhang, ``{Soft theorems in de Sitter spacetime},''
  \href{http://dx.doi.org/10.1007/JHEP01(2024)044}{{\em JHEP} {\bfseries 01}
  (2024) 044}, \href{http://arxiv.org/abs/2308.08861}{{\ttfamily
  arXiv:2308.08861 [hep-th]}}.

\bibitem{Barnich:2012aw}
G.~Barnich, A.~Gomberoff, and H.~A. Gonzalez, ``{The Flat limit of three
  dimensional asymptotically anti-de Sitter spacetimes},''
  \href{http://dx.doi.org/10.1103/PhysRevD.86.024020}{{\em Phys. Rev. D}
  {\bfseries 86} (2012) 024020},
  \href{http://arxiv.org/abs/1204.3288}{{\ttfamily arXiv:1204.3288 [gr-qc]}}.

\bibitem{Compere:2019bua}
G.~Comp\`ere, A.~Fiorucci, and R.~Ruzziconi, ``{The $\Lambda$-BMS$_4$ group of
  dS$_4$ and new boundary conditions for AdS$_4$},''
  \href{http://dx.doi.org/10.1088/1361-6382/ab3d4b}{{\em Class. Quant. Grav.}
  {\bfseries 36} no.~19, (2019) 195017},
  \href{http://arxiv.org/abs/1905.00971}{{\ttfamily arXiv:1905.00971 [gr-qc]}}.
  [Erratum: Class.Quant.Grav. 38, 229501 (2021)].

\bibitem{Compere:2020lrt}
G.~Comp\`ere, A.~Fiorucci, and R.~Ruzziconi, ``{The $\Lambda$-BMS$_4$ charge
  algebra},'' \href{http://dx.doi.org/10.1007/JHEP10(2020)205}{{\em JHEP}
  {\bfseries 10} (2020) 205}, \href{http://arxiv.org/abs/2004.10769}{{\ttfamily
  arXiv:2004.10769 [hep-th]}}.

\bibitem{deVega:1998ia}
H.~J. de~Vega, J.~Ramirez, and N.~G. Sanchez, ``{Generation of gravitational
  waves by generic sources in de Sitter space-time},''
  \href{http://dx.doi.org/10.1103/PhysRevD.60.044007}{{\em Phys. Rev. D}
  {\bfseries 60} (1999) 044007},
  \href{http://arxiv.org/abs/astro-ph/9812465}{{\ttfamily
  arXiv:astro-ph/9812465}}.

\bibitem{Ashtekar:2015lla}
A.~Ashtekar, B.~Bonga, and A.~Kesavan, ``{Asymptotics with a positive
  cosmological constant. II. Linear fields on de Sitter spacetime},''
  \href{http://dx.doi.org/10.1103/PhysRevD.92.044011}{{\em Phys. Rev. D}
  {\bfseries 92} no.~4, (2015) 044011},
  \href{http://arxiv.org/abs/1506.06152}{{\ttfamily arXiv:1506.06152 [gr-qc]}}.

\bibitem{Date:2015kma}
G.~Date and S.~J. Hoque, ``{Gravitational waves from compact sources in a de
  Sitter background},''
  \href{http://dx.doi.org/10.1103/PhysRevD.94.064039}{{\em Phys. Rev. D}
  {\bfseries 94} no.~6, (2016) 064039},
  \href{http://arxiv.org/abs/1510.07856}{{\ttfamily arXiv:1510.07856 [gr-qc]}}.

\bibitem{Chu:2016qxp}
Y.-Z. Chu, ``{Gravitational Wave Memory In dS$_{4+2n}$ and 4D Cosmology},''
  \href{http://dx.doi.org/10.1088/1361-6382/34/3/035009}{{\em Class. Quant.
  Grav.} {\bfseries 34} no.~3, (2017) 035009},
  \href{http://arxiv.org/abs/1603.00151}{{\ttfamily arXiv:1603.00151 [gr-qc]}}.

\bibitem{Compere:2023ktn}
G.~Comp\`ere, S.~J. Hoque, and E.~c. Kutluk, ``{Quadrupolar radiation in de
  Sitter: displacement memory and Bondi metric},''
  \href{http://dx.doi.org/10.1088/1361-6382/ad5826}{{\em Class. Quant. Grav.}
  {\bfseries 41} no.~15, (2024) 155006},
  \href{http://arxiv.org/abs/2309.02081}{{\ttfamily arXiv:2309.02081 [gr-qc]}}.

\bibitem{Newman:1962cia}
E.~T. Newman and T.~W.~J. Unti, ``{Behavior of Asymptotically Flat Empty
  Spaces},'' \href{http://dx.doi.org/10.1063/1.1724303}{{\em J. Math. Phys.}
  {\bfseries 3} no.~5, (1962) 891}.

\bibitem{Barnich:2011ty}
G.~Barnich and P.-H. Lambert, ``{A Note on the Newman-Unti group and the BMS
  charge algebra in terms of Newman-Penrose coefficients},''
  \href{http://dx.doi.org/10.1155/2012/197385}{{\em Adv. Math. Phys.}
  {\bfseries 2012} (2012) 197385},
  \href{http://arxiv.org/abs/1102.0589}{{\ttfamily arXiv:1102.0589 [gr-qc]}}.

\bibitem{Conde:2016rom}
E.~Conde and P.~Mao, ``{BMS Supertranslations and Not So Soft Gravitons},''
  \href{http://dx.doi.org/10.1007/JHEP05(2017)060}{{\em JHEP} {\bfseries 05}
  (2017) 060}, \href{http://arxiv.org/abs/1612.08294}{{\ttfamily
  arXiv:1612.08294 [hep-th]}}.

\bibitem{Bondi:1962px}
H.~Bondi, M.~G.~J. van~der Burg, and A.~W.~K. Metzner, ``{Gravitational waves
  in general relativity. 7. Waves from axisymmetric isolated systems},''
  \href{http://dx.doi.org/10.1098/rspa.1962.0161}{{\em Proc. Roy. Soc. Lond. A}
  {\bfseries 269} (1962) 21--52}.

\bibitem{Sachs:1962wk}
R.~K. Sachs, ``{Gravitational waves in general relativity. 8. Waves in
  asymptotically flat space-times},''
  \href{http://dx.doi.org/10.1098/rspa.1962.0206}{{\em Proc. Roy. Soc. Lond. A}
  {\bfseries 270} (1962) 103--126}.

\bibitem{Iyer:1994ys}
V.~Iyer and R.~M. Wald, ``{Some properties of Noether charge and a proposal for
  dynamical black hole entropy},''
  \href{http://dx.doi.org/10.1103/PhysRevD.50.846}{{\em Phys. Rev. D}
  {\bfseries 50} (1994) 846--864},
  \href{http://arxiv.org/abs/gr-qc/9403028}{{\ttfamily arXiv:gr-qc/9403028}}.

\bibitem{Barnich:2001jy}
G.~Barnich and F.~Brandt, ``{Covariant theory of asymptotic symmetries,
  conservation laws and central charges},''
  \href{http://dx.doi.org/10.1016/S0550-3213(02)00251-1}{{\em Nucl. Phys. B}
  {\bfseries 633} (2002) 3--82},
  \href{http://arxiv.org/abs/hep-th/0111246}{{\ttfamily arXiv:hep-th/0111246}}.

\bibitem{Strominger:2013jfa}
A.~Strominger, ``{On BMS Invariance of Gravitational Scattering},''
  \href{http://dx.doi.org/10.1007/JHEP07(2014)152}{{\em JHEP} {\bfseries 07}
  (2014) 152}, \href{http://arxiv.org/abs/1312.2229}{{\ttfamily arXiv:1312.2229
  [hep-th]}}.

\bibitem{He:2014laa}
T.~He, V.~Lysov, P.~Mitra, and A.~Strominger, ``{BMS supertranslations and
  Weinberg\textquoteright{}s soft graviton theorem},''
  \href{http://dx.doi.org/10.1007/JHEP05(2015)151}{{\em JHEP} {\bfseries 05}
  (2015) 151}, \href{http://arxiv.org/abs/1401.7026}{{\ttfamily arXiv:1401.7026
  [hep-th]}}.

\bibitem{Barnich:2010eb}
G.~Barnich and C.~Troessaert, ``{Aspects of the BMS/CFT correspondence},''
  \href{http://dx.doi.org/10.1007/JHEP05(2010)062}{{\em JHEP} {\bfseries 05}
  (2010) 062}, \href{http://arxiv.org/abs/1001.1541}{{\ttfamily arXiv:1001.1541
  [hep-th]}}.

\bibitem{Bhatkar:2022qhz}
S.~Bhatkar and D.~Jain, ``{Perturbative soft photon theorems in de Sitter
  spacetime},'' \href{http://dx.doi.org/10.1007/JHEP10(2023)055}{{\em JHEP}
  {\bfseries 10} (2023) 055}, \href{http://arxiv.org/abs/2212.14637}{{\ttfamily
  arXiv:2212.14637 [hep-th]}}.

\bibitem{Gaddam:2020mwe}
N.~Gaddam and N.~Groenenboom, ``{Soft graviton exchange and the information
  paradox},'' \href{http://dx.doi.org/10.1103/PhysRevD.109.026007}{{\em Phys.
  Rev. D} {\bfseries 109} no.~2, (2024) 026007},
  \href{http://arxiv.org/abs/2012.02355}{{\ttfamily arXiv:2012.02355
  [hep-th]}}.

\bibitem{Gaddam:2020rxb}
N.~Gaddam, N.~Groenenboom, and G.~'t~Hooft, ``{Quantum gravity on the black
  hole horizon},'' \href{http://dx.doi.org/10.1007/JHEP01(2022)023}{{\em JHEP}
  {\bfseries 01} (2022) 023}, \href{http://arxiv.org/abs/2012.02357}{{\ttfamily
  arXiv:2012.02357 [hep-th]}}.

\bibitem{Duary:2022afn}
S.~Duary, ``{AdS correction to the Faddeev-Kulish state: migrating from the
  flat peninsula},'' \href{http://dx.doi.org/10.1007/JHEP05(2023)079}{{\em
  JHEP} {\bfseries 05} (2023) 079},
  \href{http://arxiv.org/abs/2212.09509}{{\ttfamily arXiv:2212.09509
  [hep-th]}}.

\bibitem{Lysov:2014csa}
V.~Lysov, S.~Pasterski, and A.~Strominger, ``{Low\textquoteright{}s Subleading
  Soft Theorem as a Symmetry of QED},''
  \href{http://dx.doi.org/10.1103/PhysRevLett.113.111601}{{\em Phys. Rev.
  Lett.} {\bfseries 113} no.~11, (2014) 111601},
  \href{http://arxiv.org/abs/1407.3814}{{\ttfamily arXiv:1407.3814 [hep-th]}}.

\bibitem{Campiglia:2014yka}
M.~Campiglia and A.~Laddha, ``{Asymptotic symmetries and subleading soft
  graviton theorem},'' \href{http://dx.doi.org/10.1103/PhysRevD.90.124028}{{\em
  Phys. Rev. D} {\bfseries 90} no.~12, (2014) 124028},
  \href{http://arxiv.org/abs/1408.2228}{{\ttfamily arXiv:1408.2228 [hep-th]}}.

\bibitem{Campiglia:2016jdj}
M.~Campiglia and A.~Laddha, ``{Sub-subleading soft gravitons: New symmetries of
  quantum gravity?},''
  \href{http://dx.doi.org/10.1016/j.physletb.2016.11.046}{{\em Phys. Lett. B}
  {\bfseries 764} (2017) 218--221},
  \href{http://arxiv.org/abs/1605.09094}{{\ttfamily arXiv:1605.09094 [gr-qc]}}.

\bibitem{Campiglia:2016efb}
M.~Campiglia and A.~Laddha, ``{Sub-subleading soft gravitons and large
  diffeomorphisms},'' \href{http://dx.doi.org/10.1007/JHEP01(2017)036}{{\em
  JHEP} {\bfseries 01} (2017) 036},
  \href{http://arxiv.org/abs/1608.00685}{{\ttfamily arXiv:1608.00685 [gr-qc]}}.

\bibitem{Pasterski:2016qvg}
S.~Pasterski, S.-H. Shao, and A.~Strominger, ``{Flat Space Amplitudes and
  Conformal Symmetry of the Celestial Sphere},''
  \href{http://dx.doi.org/10.1103/PhysRevD.96.065026}{{\em Phys. Rev. D}
  {\bfseries 96} no.~6, (2017) 065026},
  \href{http://arxiv.org/abs/1701.00049}{{\ttfamily arXiv:1701.00049
  [hep-th]}}.

\bibitem{Pasterski:2023ikd}
S.~Pasterski, ``{A Chapter on Celestial Holography},''
  \href{http://arxiv.org/abs/2310.04932}{{\ttfamily arXiv:2310.04932
  [hep-th]}}.

\bibitem{He:2022zcf}
S.~He, P.~Mao, and X.-C. Mao, ``{$T\bar{T}$ deformed soft theorem},''
  \href{http://dx.doi.org/10.1103/PhysRevD.107.L101901}{{\em Phys. Rev. D}
  {\bfseries 107} no.~10, (2023) L101901},
  \href{http://arxiv.org/abs/2209.01953}{{\ttfamily arXiv:2209.01953
  [hep-th]}}.

\bibitem{He:2023lvk}
S.~He, P.~Mao, and X.-C. Mao, ``{Loop corrections versus marginal deformation
  in celestial holography},'' \href{http://arxiv.org/abs/2307.02743}{{\ttfamily
  arXiv:2307.02743 [hep-th]}}.

\bibitem{He:2024yzx}
S.~He and X.-C. Mao, ``{Irrelevant and marginal deformed BMS field theories},''
  \href{http://dx.doi.org/10.1007/JHEP04(2024)138}{{\em JHEP} {\bfseries 04}
  (2024) 138}, \href{http://arxiv.org/abs/2401.09991}{{\ttfamily
  arXiv:2401.09991 [hep-th]}}.

\end{thebibliography}

\end{document}